\documentclass[universe,article,submit,pdftex,moreauthors]{Definitions/mdpi} 
\usepackage{isotope}
\firstpage{1} 
\pubvolume{1}
\issuenum{1}
\articlenumber{0}
\pubyear{2023}
\copyrightyear{2023}
\datereceived{ } 
\daterevised{ } 
\dateaccepted{ } 
\datepublished{ } 
\hreflink{https://doi.org/} 

\Title{Properties of Hot Nuclear Matter}

\TitleCitation{Properties of Hot Nuclear Matter}

\Author{Omar Benhar $^{1}$, Alessandro Lovato $^{2,3}$, and Lucas Tonetto $^{4}$}

\AuthorNames{Omar Benhar, Alessandro Lovato and Lucas Tonetto}

\AuthorCitation{Benhar, O.; Lovato, A.; Tonetto , L.}

\address{%
$^{1}$ \quad INFN, Sezione di Roma, 00185 Roma, Italy; omar.benhar@roma1.infn.it\\
$^{2}$ \quad Physics Division, Argonne National Laboratory. Argonne, Illinois 60439, USA;
lovato@alcf.anl.gov \\
$^{3}$ \quad INFN, Trento Institute of Fundamental Physics and Applications, 38123 Trento, Italy\\
$^{4}$ \quad Dipartimento di Fisica, Sapienza Universit\`a di Roma, 00185 Roma, Italy ; lucas.tonetto@uniroma1.it}

\abstract{ A fully quantitative description of equilibrium and dynamical properties of hot nuclear matter will be needed for the interpretation of
the available and forthcoming astrophysical data, providing information  on
the post merger phase of a neutron star coalescence. We discuss the results of a recently developed theoretical model,
based on a phenomenological nuclear Hamiltonian including two- and three-nucleon potentials, to study the temperature dependence of
average and single-particle properties of nuclear matter relevant to astrophysical applications.
The potential of the proposed approach for describing dissipative processes leading to the appearance of bulk 
viscosity in neutron star matter is also outlined.
}

\keyword{nuclear matter; thermal effects; bulk viscosity; neutron stars.}

\begin{document}


\section{Introduction}

The interpretation of the presently available and future astronomical data providing 
information  on the post merger phase of coalescing binary neutron stars will require an accurate description of 
the properties of dense nuclear matter at temperatures as high as 100 MeV~\citep{baiotti2017,raithel2021,figura2020,figura2021,hammond2021}. Of 
great importance, in this context, will be the development of a consistent framework suitable for  modelling both the equilibrium configurations\textemdash determining the 
equation of state (EOS) of neutron star matter\textemdash and 
dissipative processes,  involving mechanisms that lead to the appearance of bulk viscosity~\cite{alford2018} and neutrino 
emission~\cite{camelio2017}.

The EOS of hot nuclear matter is often derived from dynamical models based on the independent-particle approximation, using Skyrme-type effective interactions~\cite{Prakash:1997} or the formalism of quantum field theory and the relativistic mean field (RMF) approximation~\citep{Kaplan:2014}. More comprehensive studies have been 
performed within the framework of 
Nuclear Many-Body Theory, in which the description of nuclear
dynamics is based on a phenomenological Hamiltonian strongly constrained by the observed properties of the two- and
three-nucleon systems. Calculations along this line have been carried out using both $G$-matrix 
perturbation theory and the variational approach based on the formalism of correlated wave functions
and the cluster expansion technique; see, e.g., Refs.~\cite{PhysRevC.100.054335} and~\cite{npa_cluster}.

The authors of Refs.~\cite{eos0} have developed a procedure to renormalise the coordinate-space nuclear Hamiltonian by
introducing screening effects arising from short-range nucleon-nucleon correlations. The resulting density-dependent effective potential\textemdash which includes the 
contributions of both two- and three-nucleon forces\textemdash is well-behaved, and can be employed to carry out perturbative calculations 
in the basis of eigenstates of the  non interacting system. The extension of this formalism to the case of non-zero temperature\textemdash involving a proper 
definition of the gran canonical potential needed to achieve thermodynamic consistency\textemdash is based on the assumption that at temperature 
$T \ll m_\pi$, $m_\pi \approx 150$ MeV being the mass of the $\pi$-meson, thermal effects do not significantly affect strong-interaction dynamics~\cite{benhar_2022}.

In this paper, we discuss the main features of the approach of Refs,~\cite{eos0,benhar_2022},  as well as its 
application to a variety of equilibrium and dynamical properties of nuclear matter~\cite{tonettobenhar2022}. The nuclear Hamiltonian and the 
derivation of the effective interaction are described in Section~\ref{H}, while the perturbative calculation of
the nuclear matter EOS at finite temperature is outlined in Section~\ref{pert}. Thermal effects on the single-particle 
properties of charge-neutral $\beta$-stable matter are discussed in Sections~\ref{spectra}. Finally. Section~\ref{viscosity} 
is devoted to the calculation of the bulk viscosity coefficient, driving the damping of density oscillations of neutron stars.

\section{Nuclear Hamiltonian}
\label{H}

Nuclear Many-Body Theory (NMBT) is based on the hypothesis that all nucleon systems\textemdash from the deuteron to neutron stars\textemdash  can be described in terms of point like protons and neutrons, the dynamics of which are dictated
by the Hamiltonian%
\begin{equation}
\label{hamiltonian}
H=\sum_{i}\frac{p_i^2}{2m} + \sum_{i<j}v_{ij}+\sum_{i<j<k}V_{ijk} \ ,
\end{equation}
with $m$ and ${\bf p}_i$ being the mass and momentum of the $i$-th particle.\footnote{In this article, we adopt the
system of natural units, in which $\hbar=c=k_B=1$, and, unless
otherwise specified, neglect the small proton-neutron mass difference.}

Nucleon-nucleon (NN) potentials that are local or semi-local in coordinate space are usually written in
the form
\begin{equation}
    v_{ij} = \sum_p v^p (r_{ij}) O_{ij}^p \ ,
    \label{eq:vij}
\end{equation}
where $r_{ij} = |{\bf r}_i - {\bf r}_j|$ is the distance between the interacting particles. They are designed
to reproduce the measured properties of the two-nucleon system, in both bound and scattering states, and reduce
to the Yukawa one-pion exchange potential at large distances. The sum in Eq.~\eqref{eq:vij} includes up
to eighteen terms, with the corresponding operators, $O^p$, being required to describe the strong spin-isospin
dependence and non central nature of nuclear forces ($i=1,\ldots,6$), as well as the occurrence of spin-orbit and other angular-momentum dependent interactions ($i=7,\ldots 14$). Hihgly accurate phenomenological potentials, such as the Argonne $v_{18}$ (AV18) model, also feature additional terms accounting for small violations of charge symmetry and charge independence ($i=15,\ldots,18$)~\cite{Wiringa:1994wb}.

The addition of the three-nucleon (NNN) potential $V_{ijk}$ is needed to model the effects of {\it irreducible} three-body
interactions, reflecting the appearance of processes involving the internal structure of the nucleons.
Nuclear Hamiltonians comprising the AV18 NN potential and a phenomenological NNN potential designed to explain the binding 
energies of \isotope[3][]{He} and \isotope[4][]{He} and the empirical equilibrium density of isospin-symmetric matter\textemdash such as the
widely used Urbana IX (UIX) model~\cite{UIX_2,Pudliner:1995wk}\textemdash have been shown to possess a remarkable predictive power. The results of Quantum Monte Carlo (QMC) calculations, extensively
reviewed in Ref.\cite{QMC}, demostrate that the AV18 + UIX Hamiltonian is capable describe the energies of the ground and 
low-lying excited states of nuclei with mass number $A\leq 8$ to few percent accuracy.  
 
\subsection{Renormalisation of the nucleon-nucleon interaction}

Owing to the presence of a strong repulsive core,
the matrix elements of the NN potential between eigenstates of the non interacting system are large, and standard perturbation theory cannot be used to carry out calculations of nuclear-matter properties.

The renormalisation scheme based on the formalism of correlated basis functions (CBF) and the cluster expansion technique~\cite{CLARK197989,FF:CBF} allows one to 
determine an effective interaction suitable to carry out perturbative calculations in coordinate space. This approach, originally proposed by Cowell and Pandharipande in the early 2000s~\cite{cowell:2003,cowell:2004} and further developed by the authors of Refs.~\cite{bv,Lovato:2012ux,LBGL}, has been extensively employed to study nuclear matter and neutron stars~\cite{eos0,benhar_2022,tonettobenhar2022}.
The resulting potential, can be written as in Eq.~\eqref{eq:vij} including terms with $i \leq 6$, associated with the operators
\begin{align}
O^{p \leq 6}_{ij} = [1, (\boldsymbol{\sigma}_{i}\cdot\boldsymbol{\sigma}_{j}), S_{ij}]
\otimes[1,(\boldsymbol{\tau}_{i}\cdot\boldsymbol{\tau}_{j})]  \ .
\label{av18:2}
\end{align}
Here $\boldsymbol{\sigma}_{i}$ and $\boldsymbol{\tau}_{i}$ are Pauli matrices acting in spin and isospin space, respectively, while the angular dependence is described by the tensor operator $S_{ij}$, defined as
\begin{align}
S_{ij}=\frac{3}{r_{ij}^2}
(\boldsymbol{\sigma}_{i}\cdot{\bf r}_{ij}) (\boldsymbol{\sigma}_{j}\cdot{\bf r}_{ij})
 - (\boldsymbol{\sigma}_{i}\cdot\boldsymbol{\sigma}_{j}) \ .
 \label{S12}
\end{align}
Note that the OPE potential can also be written in terms of the $O^{p \leq 6}_{ij}$ defined by  Eqs.\eqref{av18:2} and \eqref{S12}.

The CBF effective interaction is {\em defined} by the equation
\begin{equation}
\label{def:veff}
\langle H \rangle = \langle\Psi_0 | H | \Psi_0 \rangle = T_F + \langle \Phi_0 | \sum_{i<j} v_{ij}^\text{eff} | \Phi_0\rangle \ .
\end{equation}
where, $| \Phi_0\rangle$ and $T_F$ denote the ground state of the non interacting Fermi gas at density $\varrho$ and the
corresponding energy, respectively, while $H$ is the nuclear Hamiltonian of Eq.~\eqref{hamiltonian}.
The {\it correlated} ground state, $| \Psi_0\rangle$, is obtained from the corresponding Fermi gas state $| \Phi_0\rangle$ through the transformation
\begin{equation}
\label{def:corrfun}
|\Psi_0 \rangle \equiv \frac{{F}|\Phi_0\rangle}{\langle \Phi_0 | {F}^\dagger {F} |\Phi_0\rangle^{{1/2}}} \ ,
\end{equation}
where the operator ${F}$
is a symmetrized product of two-body correlation operators, whose structure is chosen in such a way as to  reflect the complexity of NN forces.

The effective interaction employed to obtain the results discussed in this paper has been derived following the 
procedure described in Ref.~\cite{Lovato:2012ux}, which allows to include the 
contribution of three-nucleon clusters to the ground-state expectation value $\langle H \rangle$ appearing in the 
left-hand side of Eq.~\eqref{def:veff}. This feature is essential to take into account three-nucleon interactions, which play a dominant role in the high-density  regime relevant to astrophysical applications.

\subsection{The CBF effective interaction}
\label{CBF:veff}

The nuclear Hamiltonian employed to obtain the CBF effective interaction  consists of the Argonne $v_6^\prime$ (AV6P) NN potential\textemdash determined projecting the full AV18 potential onto the operator basis of Eq.\eqref{av18:2}~\cite{Wiringa:2002ja}\textemdash and the UIX NNN potential~\cite{Pudliner:1995wk}.
The AV6P potential predicts the binding energy and electric quadrupole
moment of the deuteron with accuracy of 1\%, and 4\%, respectively, and provides an excellent fit of the elastic NN scattering phase
shifts in the $^1{\rm S}_0$ channel\textemdash which is dominant in pure neutron matter\textemdash up to lab energy $\sim600$ MeV, well above pion production threshold.. 

The UIX potential is written in the form
\begin{align}
V_{ijk}=V_{ijk}^{2\pi}+V_{ijk}^{R} \ ,
\end{align}
where the first term is the attractive Fujita-Miyazawa potential~\cite{Fujita:1957zz}\textemdash describing two-pion exchange NNN interactions
with excitation of a $\Delta$-resonance in the intermediate state\textemdash while the purely phenomenological 
$V_{ijk}^{R}$ is an isoscalar repulsive term, the strength of which is fixed in such a way as to reproduce the saturation density of isospin symmetric matter inferred from nuclear systematics~\cite{UIX_2,Pudliner:1995wk}.

\begin{figure}[H]
\includegraphics[width=8.0 cm]{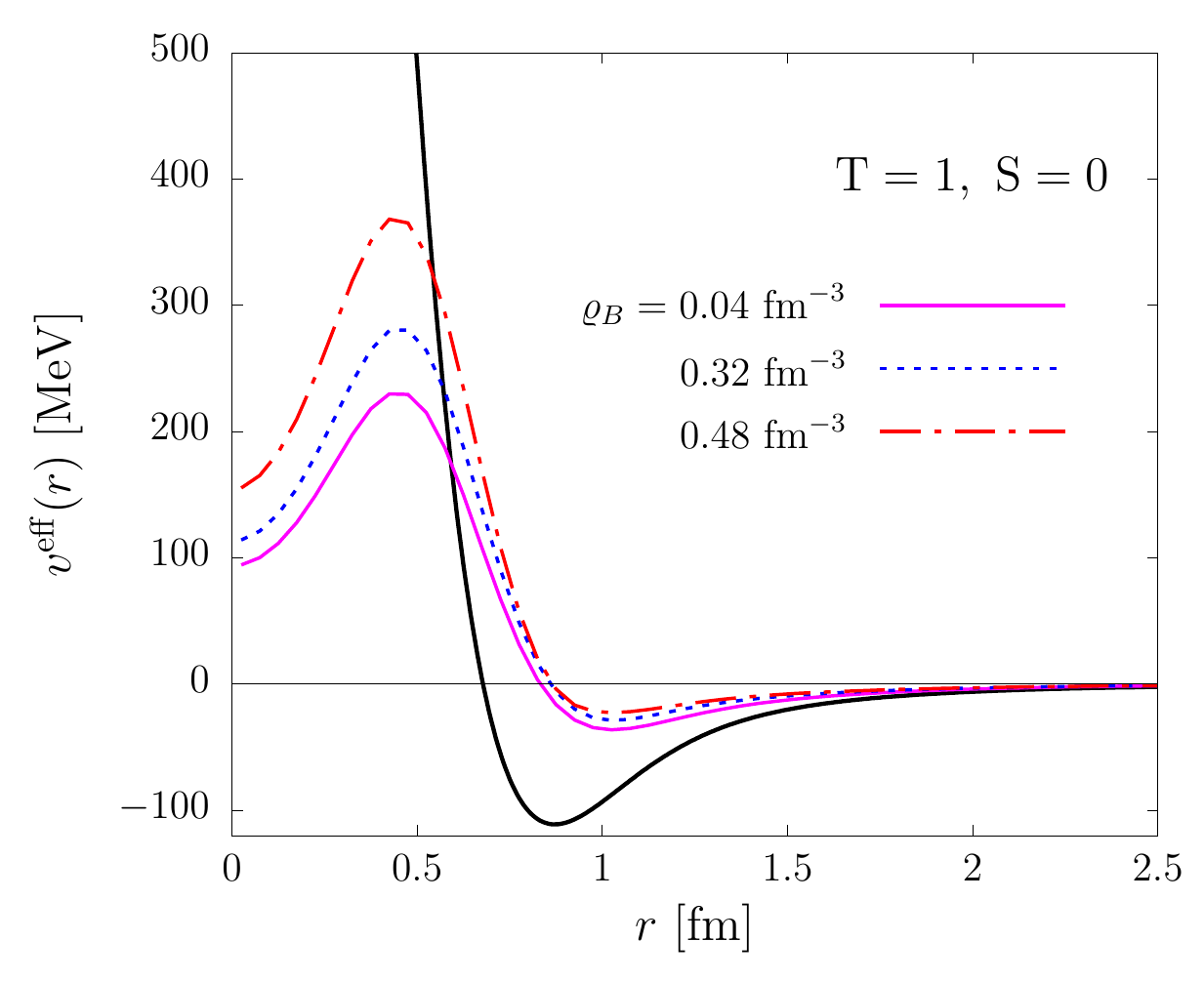}
\caption{Radial dependence of the CBF effective potential  in the  $S=0$, $T=1$  channel.
The solid, dashed, and dot-dash lines correspond to baryon number density $\varrho =$ 0.04, 0.32 and 0.48 ${\rm fm}^{-3}$.
For comparison, the thick solid line shows the bare AV6P potential. \label{densdep}}
\end{figure}

Note that the CBF effective interaction depends on density through both dynamical correlations, described by the operator $F$ of Eq.\eqref{def:corrfun}, and statistical correlations, arising from the antisymmetric nature
 of the state $ | \Phi_0\rangle$. The radial dependence of $v^{\rm eff}$ in the $S=0$ and $T=1$ channel at baryon density $\varrho =$ 0.04, 0.32 and 0.48 fm$^{-3}$ is displayed in Fig.~\ref{densdep}. The effect of renormalisation clearly emerges from the comparison with the bare AV6P potential.

Recent studies of the EOS of cold neutron matter\textemdash performed by the authors of Ref.~\cite{Lovato:2022} using accurate computational techniques\textemdash show that the
predictions of the somewhat simplified AV6P + UIX Hamiltonian are very close to those obtained from the full AV18 + UIX model employed by Akmal, Pandharipande and Ravenhall~\cite{akmal:1998}.

\section{Many-Body Perturbation Theory at Finite Temperature}
\label{pert}

Let us consider, for simplicity, a one-component Fermi system. The derivation of perturbation theory at finite-temperature 
is based on the solution of the Bloch equation 
\begin{align}
\label{bloch}
- \frac { \partial \Phi }{ \partial \beta } = (H - \mu N)\Phi \ ,
\end{align}
where
\begin{align}
\label{def:Phi}
\Phi(\beta) = e^{-\beta(H - \mu N)} \ , 
\end{align}
with the initial condition $\Phi(0) = 1$; see, e.g., Ref.~\cite{Thouless}.
In the above equations $\beta = 1/T$, while $H$ and $\mu$ denote the Hamiltonian and the chemical potential, respectively.

The perturbative expansion of the grand canonical  partition function $Z = {\rm Tr} \ \Phi$ is easily obtained
exploiting the formal similarity between Eq.~\eqref{bloch} and the time-dependent Schr\"odinger equation of quantum mechanics,
and rewriting the Hamiltonian in the form
\begin{align}
\label{def:H}
H & = H_0 + H_I .
\end{align}
Substitution of Eq.~\eqref{def:H} into the right-hand side of the Bloch equation, leading to
\begin{align}
\nonumber
- \frac { \partial \Phi }{ \partial \beta }  = [ (H_0 - \mu N) + H_I ]\Phi  = (H_0^\prime + H_I ) \Phi  \ ,
\end{align}
shows that the formalism of time-dependent perturbation theory can be readily generalised by replacing
$t \to -i \beta$, and using $H_0^\prime$ to define operators in the interaction picture.

The fundamental relation
\begin{align}
\label{fundamental}
\Omega = -\frac{1}{\beta} \ln Z = -PV = F - \mu N = E - TS - \mu N  \ ,
\end{align}
where $V$ is the normalization volume, provides the link between the grand canonical potential $\Omega$, the pressure $P$, and the free energy $F = E - TS$, with $E$ and $S$ being the energy and  entropy
of the system, respectively; see, e.g., Ref.~\cite{landau}. From Eq.~\eqref{fundamental}
if follows that
\begin{align}
P = - \frac{\Omega}{V} \ \ \ , \ \ \ S = - \frac{\partial \Omega}{\partial T} \ \ \ , \ \ \  N = - \frac{\partial \Omega}{\partial \mu} \ . 
\end{align}
In the following, we will discuss the application of the above results to a system described by the Hamiltonian
\eqref{def:H}, with
\begin{align}
\label{def:H0}
H_0 = \sum_k e_k a^\dagger_k a_k \  ,
\end{align}
where, in general
\begin{align}
\label{def:ek}
e_k = \frac{ {\bf k}^2}{2m} + U_k = t_k + U_k \ ,
\end{align}
and
\begin{align}
\label{def:HI}
H_I = \sum_{ k, k^\prime, q, q^\prime }  \langle k^\prime q^\prime | v |  k q \rangle
 a^\dagger_{k^\prime} a^\dagger_{q^\prime} a_{q} a_{k}
- \sum_k U_k a^\dagger_k a_k \  .
\end{align}
Here, the label $k$ specifies both the particle momentum and the discrete quantum numbers corresponding to one-particle states,
$a^\dagger_{k}$ and $a_{k}$ denote creation and annihilation operators, respectively, and $v$ is the
potential describing interparticle  interactions. The single-particle potential $U_k$, which in principle does not
affect the results of calculations of physical quantities, is  chosen in such a way as to improve the convergence of
the perturbative expansion, or fulfil specific conditions; see, e.g., Ref.~\cite{Baldo}.

It has to be pointed out that, according to Eq.~\eqref{fundamental}, the pressure can be written in the form
\begin{align}
\label{HV}
P = \varrho \Big( \mu - \frac{F}{N} \Big) \ ,
\end{align}
with $\varrho = N/V$, implying that at equilibrium, that is, for $P=0$,  $\mu = F/N$.
This result is the generalisation of the
Hugenholtz-Van Hove theorem~\citep{HVH} to the case of non vanishing temperature.

It should be emphasised that, when used in conjunction with the CBF effective interaction discussed in Section~\ref{CBF:veff}, the 
perturbative approach based on the Hamiltonian defined by Eqs.~\eqref{def:H}~and~\eqref{def:H0}-\eqref{def:HI} allows 
to take into account two- and three-nucleon interactions in a fully consistent fashion.

\subsection{Perturbative expansion}
\label{sec:energy}

At first order in $H_I$, the grand canonical potential is given by~\cite{lejeune:NPA}
\begin{align}
\label{Omega:pert}
\Omega = \Omega _0 + \Omega_1 \ ,
\end{align}
with
\begin{align}
\label{Omega0}
\Omega_0 & = - \frac{1}{\beta} \sum_k \ln \big\{ 1 + e^{-[\beta(e_k - \mu)]} \big\} \ , \\
\label{Omega1}
\Omega_1 & = \frac{1}{2} \sum_{k k^\prime}  \langle k k^\prime | v | k k^\prime \rangle_A \ n_k n_{k^\prime}
 - \sum_k U_k n_k \ ,
\end{align}
where $ | k k^\prime \rangle_A = | k k^\prime \rangle - | k^\prime k \rangle$ denotes an antisymmetrised two-particle state,
and $n_k$ is the Fermi distribution, defined as
\begin{align}
n_k = \big[ 1 + e^{\beta(e_k - \mu)} \big]^{-1} \ .
\label{fermidist:1}
\end{align}
From Eqs.~\eqref{Omega0} and \eqref{Omega1} it follows that the free energy per particle
\begin{align}
\label{F1}
\frac{F}{N} = \frac{1}{N} ( \Omega_0 + \Omega_1) + \mu   \ ,
\end{align}
can be cast in the form
\begin{align}
\label{F2}
\frac{F}{N}  & = \frac{1}{N} \Big\{ \sum_k  t_k n_k + \frac{1}{2} \sum_{k, k^\prime} \langle k k^\prime | v | k k^\prime \rangle_A \ n_k n_{k^\prime} \\ 
 & + \frac{1}{\beta} \sum_k \big[ n_k \ln n_k + (1-n_k) \ln (1-n_k) \big]
 \nonumber
+ \mu \Big(1 - \frac{1}{N} \sum_k n_k \Big) \Big\} \ .
\end{align}
In principle, for any assigned values of temperature  and chemical potential, the above equations provide a scheme for the determination of the equation of
state of nuclear matter at finite temperature,  $P = P(\mu,T)$. Because baryon number is conserved by all known interactions, however, in nuclear matter it is more convenient to use baryon density as an independent variable, and determine the chemical
potential from the relation
\begin{align}
\varrho = - \frac{1}{V} \frac{\partial}{\partial \mu} \big( \Omega _0 + \Omega_1 \big) \ .
\end{align}
In the $T \to 0$ limit the momentum distribution reduces to the Heaviside step function $\theta(\mu - e_k)$, and
the chemical potential is given by  $\mu = e_{k_F}$, with
the Fermi momentum defined as $k_F = \big( 6 \pi^2 \varrho / \nu \big)^{1/3}$.

\subsection{Thermodynamic consistency}

For $T\neq0$ and density-dependent potentials, thermodynamic consistency is not trivially achieved 
at any given order of perturbation theory. A clear manifestation of this difficulty is the mismatch between the value of pressure
obtained from Eq.~\eqref{HV} and the one resulting from the alternative\textemdash although in principle equivalent\textemdash thermodynamic expression
\begin{align}
P= - \frac{\partial F}{\partial V} = \varrho^2 \frac{\partial}{\partial \varrho} \frac{F}{N} \ .
\end{align}

A procedure fulfilling the requirement of thermodynamic consistency by construction can be derived from
a variational approach, based on minimisation of the trial grand canonical potential~\citep{heyer:PLB}
\begin{align}
\widetilde{\Omega} =  \sum_k  t_k n_k & + \frac{1}{2} \sum_{k, k^\prime}
 \langle k k^\prime | v | k k^\prime \rangle_A \ n_k n_{k^\prime} \\
 \nonumber
 & + \frac{1}{\beta} \sum_k \big[ n_k \ln n_k + (1-n_k) \ln (1-n_k) \big] ,
\end{align}
with respect to the form of distribution $n_k$.  Note that the above
expression\textemdash the use of which is fully legitimate in the variational context\textemdash can also be obtained in first order perturbation theory neglecting terms involving $\partial \Omega_1 / \partial T$ and
$\partial \Omega_1 / \partial \mu$~\citep{lejeune:NPA}.

The condition
\begin{align}
\frac{\delta \widetilde{\Omega}  }{\delta n_k} = 0  \ ,
\end{align}
turns out to be satisfied by the distribution function
\begin{align}
\label{Fermi:consistent}
n_k = \big\{ 1 + e^{\beta[(t_k + U_k + \delta e) - \mu]} \big\}^{-1} \ ,
\end{align}
with
\begin{align}
\label{HF}
U_k = \sum_{k^\prime}
 \langle k k^\prime | v | k k^\prime \rangle_A \ n_{k^\prime} \ ,
\end{align}
and
\begin{align}
\label{def:deltae}
\delta e= \frac{1}{2} \sum_{k, k^\prime} \langle k k^\prime | \frac{ \partial v}{\partial \varrho}  | k k^\prime \rangle_A \ n_k n_{k^\prime} \ .
\end{align}
Within the above  scheme, that reduces to the standard Hartee-Fock approximation in the case of density-independent potentials, all thermodynamic functions
at given temperature and baryon density can be consistently obtained using the distribution $n_k$ of Eq.~\eqref{Fermi:consistent}. Note, however, that, because both $U_k$ and $\delta e$
depend on $n_k$, see Eqs.~\eqref{HF} and \eqref{def:deltae}, calculations must be carried out self-consistently, applying an iterative procedure.

\section{Equilibrium properties of hot nuclear matter}
Consider now a nucleon system at temperature $T$, baryon number density $\varrho$ and proton number density $\varrho_p = Y_p ~\varrho$.  
At first order in the CBF effective interaction $v^{\rm eff}$, the internal energy per nucleon can be written in the form~\cite{tonettobenhar2022}
\begin{align}
\label{int:en}
\frac{E}{N} = \frac{1}{N} \Big\{ \sum_{ \alpha {\bf k} }   \ \frac{ {\bf k}^2 }{2m}  \ n_\alpha(k,T)
+ \frac{1}{2} \sum_{ \alpha {\bf k} } \sum_{ \alpha^\prime {\bf k}^\prime }
\langle \alpha {k} , \alpha^\prime  {k}^\prime | v^{\rm eff} | \alpha {k} , \alpha^\prime  {k}^\prime \rangle_A
\ n_\alpha(k,T) n_{\alpha^\prime}(k^\prime,T) \Big\} \ .
\end{align}
In the above equations, the index $\alpha = n, p$ labels neutrons and protons, respectively, 
${\bf k}$ is the nucleon momentum, $k = |{\bf k}|$,
and $| \alpha {k} , \alpha^\prime  {k}^\prime \rangle_A$ denotes an antisymmetrised two-nucleon state. Conservation of baryon number
obviously implies that $Y_n = 1 - Y_p$.

The temperature dependence is
described by the Fermi distributions 
\begin{align}
\label{fermidist}
n_\alpha(k,T)  = \Big\{ 1 +  \exp { [ \beta ( e_{\alpha k} - \mu_\alpha ) ] } \Big\}^{-1}  \ .
\end{align}
where the single-particle energies are defined as,
\begin{align}
\label{ek}
e_{\alpha k} = e^{\rm HF}_{\alpha k} + \delta e \ ,
\end{align}
with
\begin{align}
\label{eHFk}
e^{\rm HF}_{\alpha k} =\frac{{\bf k}^2}{2m} & + \sum_{ \alpha^\prime {\bf k}^\prime }
\langle \alpha {k} , \alpha^\prime  {k}^\prime | v^{\rm eff} | \alpha {k} , \alpha^\prime  {k}^\prime \rangle_A  \ n_\alpha(k^\prime,T) \ ,
\end{align}
and
\begin{align}
 \delta e = \frac{\varrho}{2} \sum_{ \alpha {\bf k} } \sum_{ \alpha^\prime {\bf k}^\prime } 
 \langle \alpha {k} , \alpha^\prime  {k}^\prime | \frac{\partial v^{\rm eff}}{\partial \varrho} | \alpha {k} , \alpha^\prime  {k}^\prime \rangle_A 
\label{deltae}
 \ n_\alpha(k,T) n_{\alpha^\prime}(k^\prime,T) ,
\end{align}
The correction to the Hartree Fock (HF) spectrum is needed to satisfy
the requirement of thermodynamic consistence, and vanishes in the case of a density-independent potential~\cite{benhar_2022}. The chemical potentials $\mu_\alpha$ are determined by the normalisation conditions
\begin{align}
\label{def:chempot}
2  \ \sum_{ \alpha {\bf k} } n_\alpha(k,T)  = N_\alpha \ ,
\end{align}
where $N_\alpha$ denotes the number of  particles of species $\alpha$, the fractional density of which is 
$Y_\alpha = N_\alpha / N_B = \varrho_\alpha/\varrho$. 
Note that the above definitions
imply that both the single-nucleon energies and the chemical potentials depend on temperature through the Fermi distribution.

The free energy per nucleon is obtained from
\begin{align}
\label{def:F}
\frac{F}{N} = \frac{1}{N}  \big( E - TS \big) \ , 
\end{align}
with the internal energy per nucleon of Eq.~\eqref{int:en} and the entropy per nucleon defined as
\begin{align}
\label{def:S}
\frac{S}{N}   = - \sum_{ \alpha {\bf k} }
\Big\{  n_\alpha(k,T) \ln {n_\alpha(k,T)} 
 +  \big[ 1 - n_\alpha(k,T) \big]  \ln{ \big[ 1 - n_\alpha(k,T) \big] } \Big\} \ .
\end{align} 
\begin{figure}[H]
\includegraphics[width=6.9 cm]{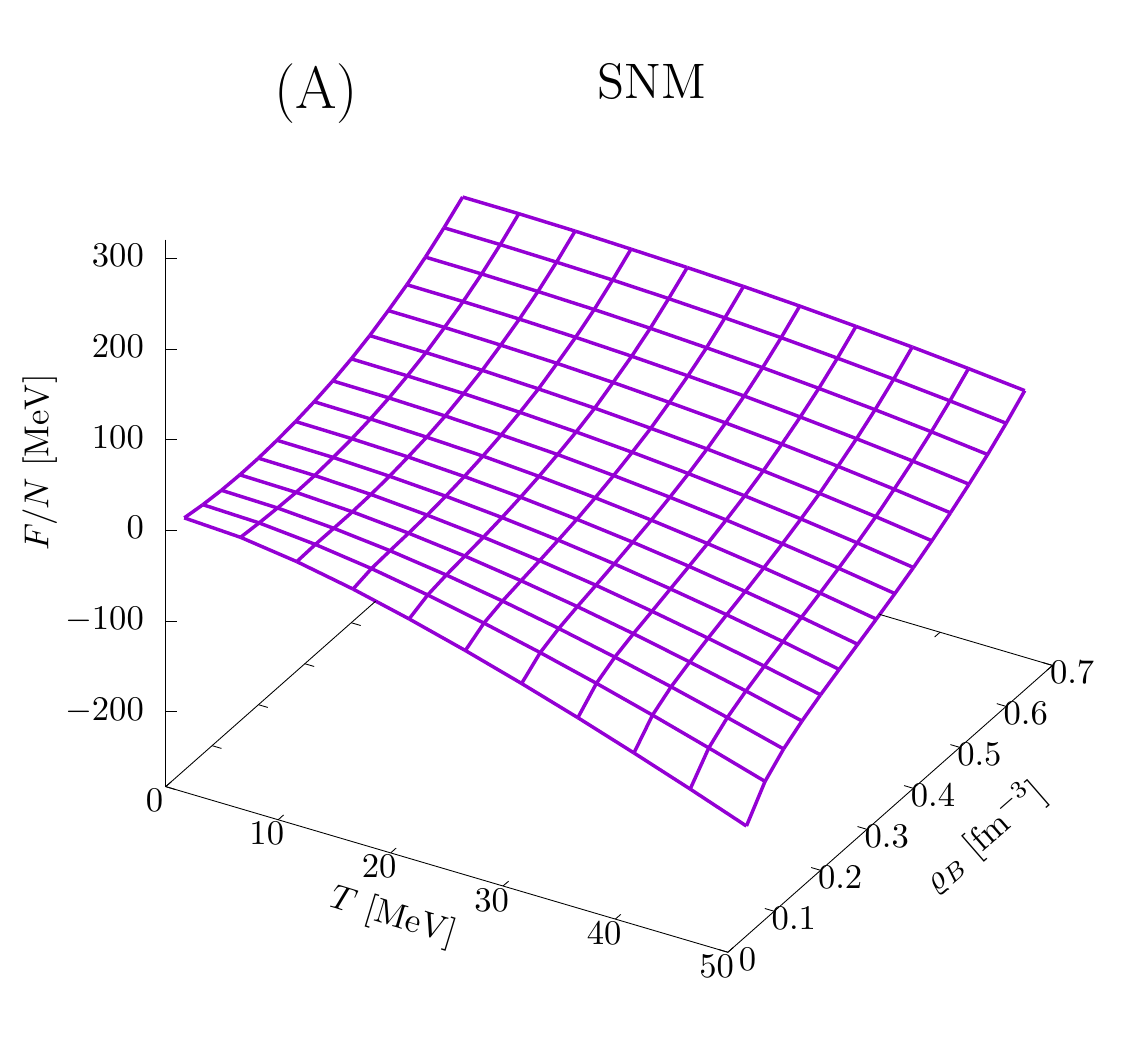}\includegraphics[width=6.9 cm]{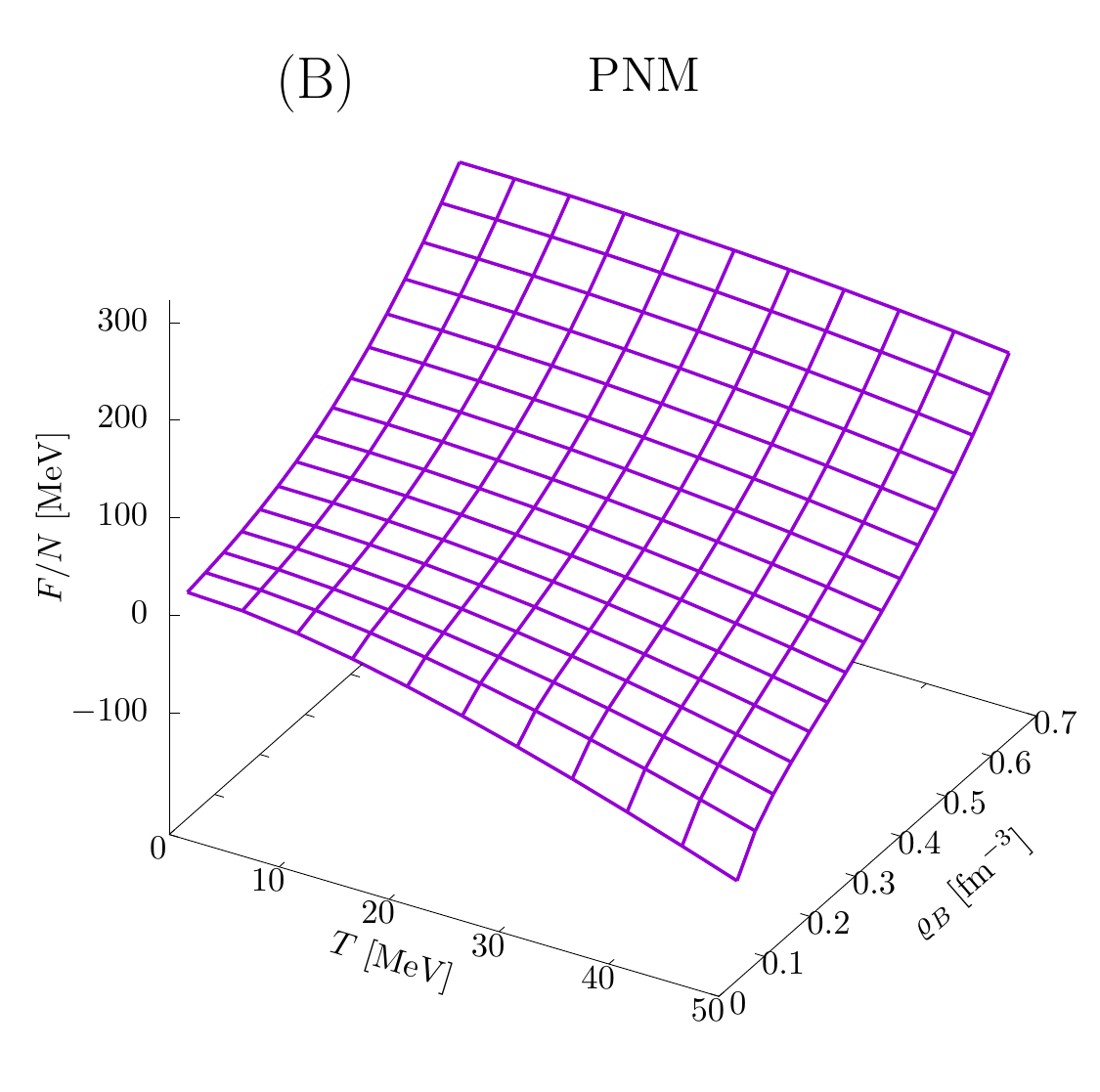}
\caption{Density and temperature dependence of the free energy per nucleon of SNM (A) and PNM (B), computed using Eqs.~\eqref{int:en}-\eqref{def:S}, with  the CBF effective interaction discussed in Section~\ref{CBF:veff}. \label{SNM_PNM_3D}}
\end{figure}   
Figure~\ref{SNM_PNM_3D} shows the density and temperature dependence of the free energy per
nucleon of isospin-symmetric matter (SNM) and pure neutron matter (PNM), corresponding to proton fraction $Y_p = 0.5 $ and 0, respectively, obtained from the  procedure described above using the CBF effective interaction.

\section{Thermal effects on nuclear matter properties}
\label{spectra}

In the temperature regime discussed in this paper, thermal modifications of nuclear matter properties
arise primarily from the Fermi distributions, defined by Eq.~\eqref{fermidist}. Comparison to the $T \to 0$ limit shows that the probability distribution $n_\alpha(k,T>0)$ is reduced from unity in the region corresponding to $\mu_{\alpha}  -T \lesssim  e{_{\alpha k}} \lesssim \mu_{\alpha}$, and acquires non vanishing positive values for $\mu_{\alpha}  \lesssim e{_{\alpha k}} \lesssim  \mu_{\alpha} + T$. It follows that, for any given temperature $T$, the extent of thermal modifications to the Fermi distribution is driven by the ratio $2T/\mu_{\alpha}$. This observation in turn implies that, because the chemical potential is a monotonically increasing function of  the particle density $\varrho_\alpha$ over a broad range of temperatures, for any given $T$ thermal effects turn out to be  more significant at lower $\varrho_\alpha$. On the other hand,  they become vanishingly small in the high-density regime, in which degeneracy  dominates. 

\subsection{Charge-neutral $\beta$-stable matter at finite temperature}

In charge-neutral matter consisting of neutrons, protons and leptons in equilibrium with respect to the weak interaction processes
\begin{align}
n \to p + \ell + {\bar \nu}_\ell \ \ \ \ , \ \ \ \ p + \ell^- \to n + \nu_\ell \ ,
\end{align}
where $\ell = e, \mu$ labels the lepton flavour, the proton fraction $Y_p$ is uniquely determined by the equations
\begin{align}
\label{beta:eq1}
\mu_n - \mu_p = \mu_{\ell} \ \ \  , \ \ \  Y_p = \sum_\ell Y_\ell \ .
\end{align}
At densities such that the electron chemical potential does not exceed the rest mass of the muon, $m_\mu =~105.7$~MeV, the sum appearing in the
above equation includes electrons only. At higher densities\textemdash typically at $\varrho \gtrsim \varrho_0$, with $\varrho_0 = 0.16 \ {\rm fm}^{-3}$ being 
the baryon number density 
of isospin-symmetric matter in thermodynamic equlibrium at $T=0$\textemdash  the appearance of muons becomes energetically favoured, and must be taken into account.

\begin{figure}[htb]
\includegraphics[scale=0.6]{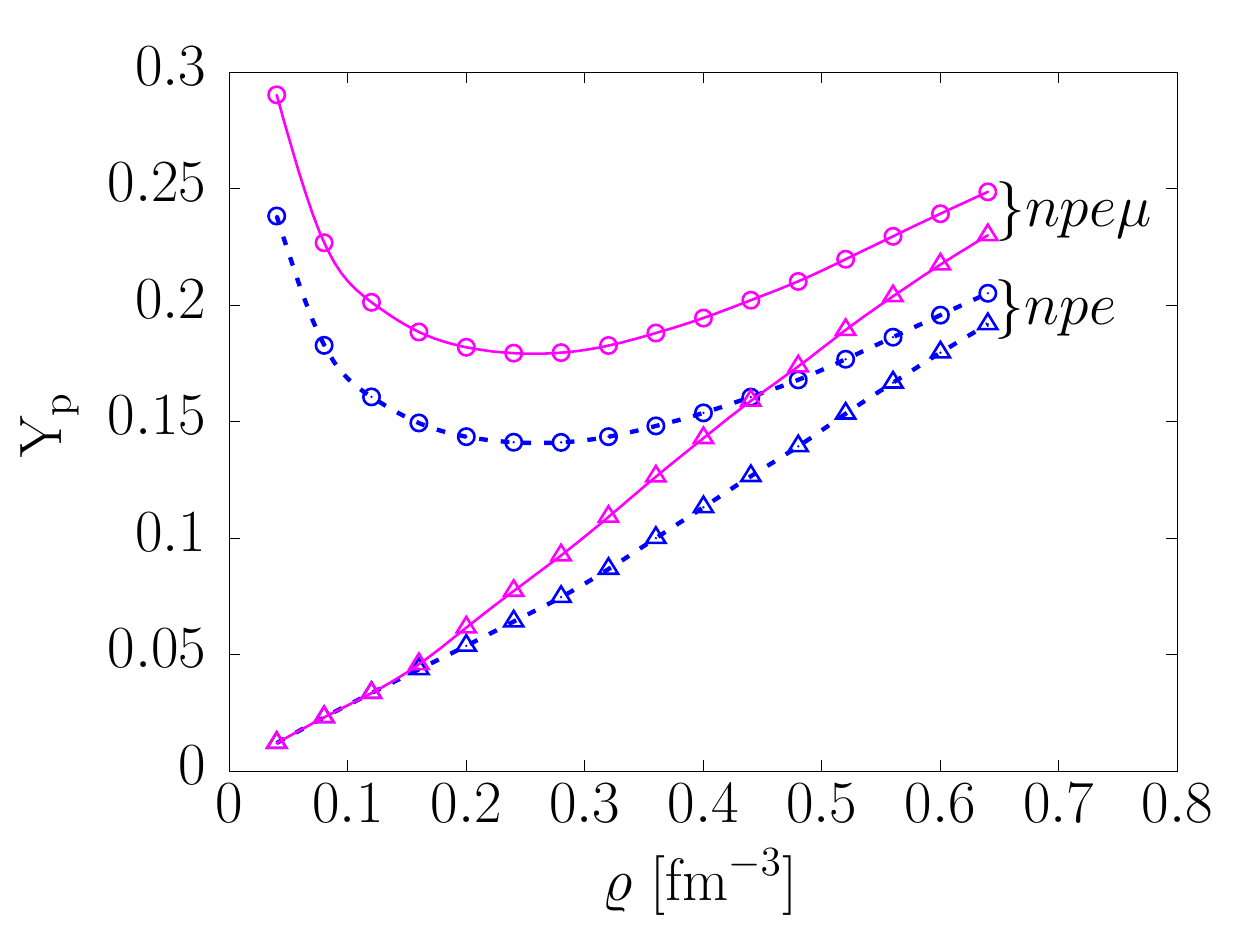}
\vspace{-0.25cm}
     \caption{Density dependence of the proton fraction in charge-neutral $\beta$-stable matter. Solid lines marked with triangles and circles correspond to $npe\mu$ matter at $T=$ 0 amd 50 MeV, respectively.
     The same quantities in $npe$ matter are represented by dashed lines. From Ref.~\cite{tonettobenhar2022}.}
\label{prot:frac}
\end{figure}

The solid lines of Fig.~\ref{prot:frac} show the density dependence of the proton fractions corresponding to $\beta$-equilibrium of matter consisting of protons, neutrons, electrons and muons, or $npe\mu$ matter, at $T=$ 0 (triangles) and 50 MeV (circles)~\cite{tonettobenhar2022}. The results have been obtained 
assuming that neutrinos do not interact with matter, and have therefore vanishing chemical potential. For comparison, the same quantities in $npe$ matter, in which the muon contribution is not included, are displayed
by the dashed lines.
The most prominent thermal effect is a significant departure from the monotonic behaviour observed in cold matter. The emergence of a minimum in the
density dependence of the proton fraction results from the balance between the thermal and degeneracy contributions to the  chemical potentials appearing in Eq.~\eqref{beta:eq1}.
For $T \gtrsim 20$~MeV and low density, typically $\varrho \lesssim \varrho_0$, the thermal contribution\textemdash whose leading order term can be written in the form $\delta \mu_\alpha \propto T^2/\varrho_\alpha^{1/3}$\textemdash turns out to be much larger for protons than for neutrons, and $\beta$-equilibrium requires large proton fractions. 
The results displayed in Fig.~\ref{prot:frac}, showing that $Y_p$ does not exceed 25\% for $\varrho_0/2 \leq \varrho \leq  4 \varrho_0$, imply that in 
$\beta$-stable matter thermal effects affect mainly the proton distributions.

The description of single-particle dynamics in interacting many-body systems is largely based on the use of the effective 
mass $m^\star_\alpha$, defined as 
\begin{align}
\label{def:mstar}
\frac{1}{m^\star_\alpha} = \left( \frac{1}{k} \frac{ d e_{\alpha k} }{d k} \right)_{k = {k_{F_\alpha}}} \ , 
\end{align}
with $e_{\alpha k}$ given by Eq.\eqref{ek}. The effective mass dictates the nucleon dispersion relation in matter, 
which plays a critical role in determining the rates of many processes relevant 
to neutron star properties, such as the bulk viscosity to be discussed in Section~\ref{viscosity}. 
The momentum dependence of the nucleon spectra in cold nuclear matter is often parametrised according to~\cite{camelio2017}
\begin{align}
\label{ek:quad}
e_{\alpha k} = \frac{k^2}{2 m^\star_0} + U_\alpha \ ,
\end{align}
where $m^\star_0$ denotes the value of  ${m_\alpha^\star}$  at $T=0$, while the offset $U_\alpha$ from the free-nucleon 
spectrum\textemdash the value of which depends on {\it both} temperature and density\textemdash is determined by the requirement that the above approximation reproduce the results of the full microscopic calculation at $k=0$.

\begin{figure}[h]
\vspace{0.5cm}
\includegraphics[scale=0.5250]{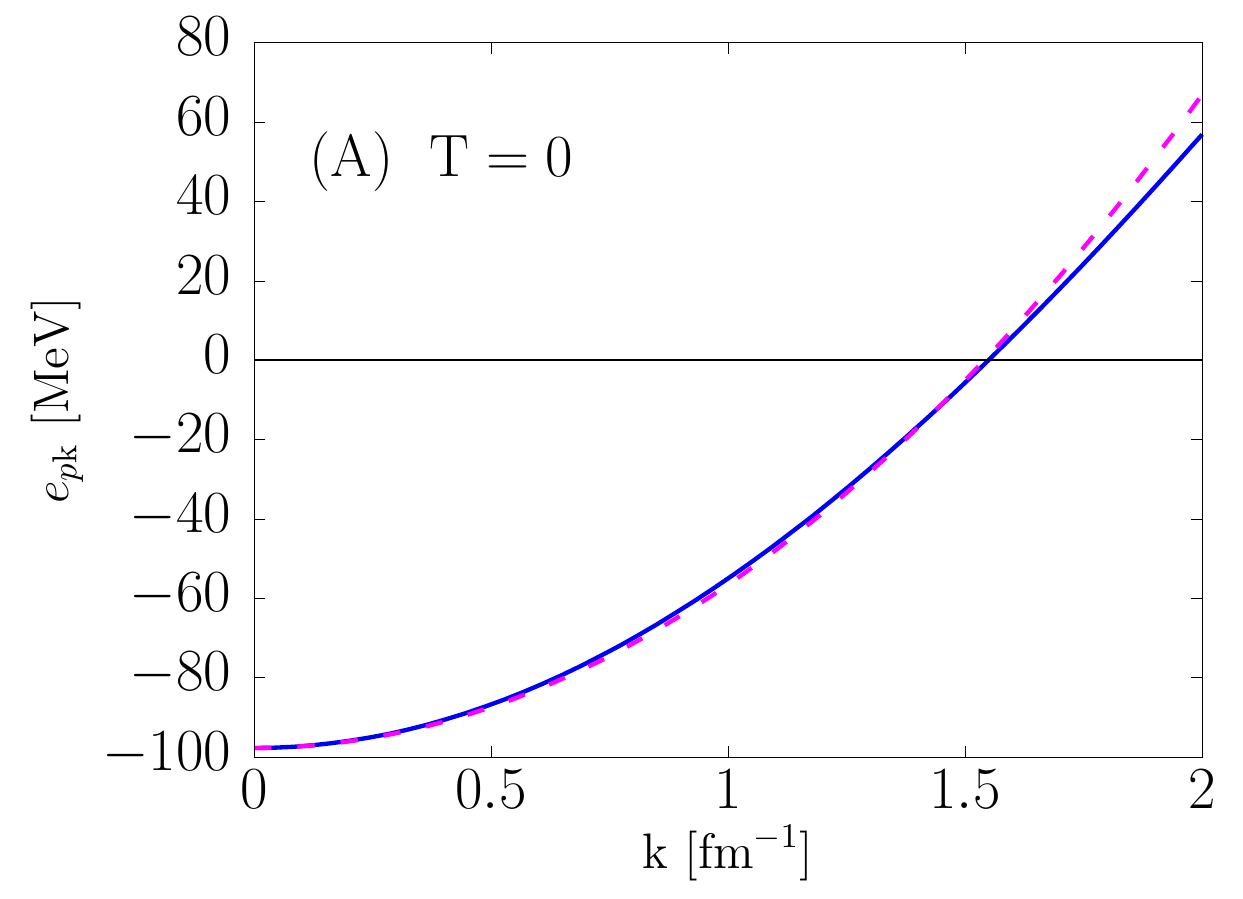}\includegraphics[scale=0.5250]{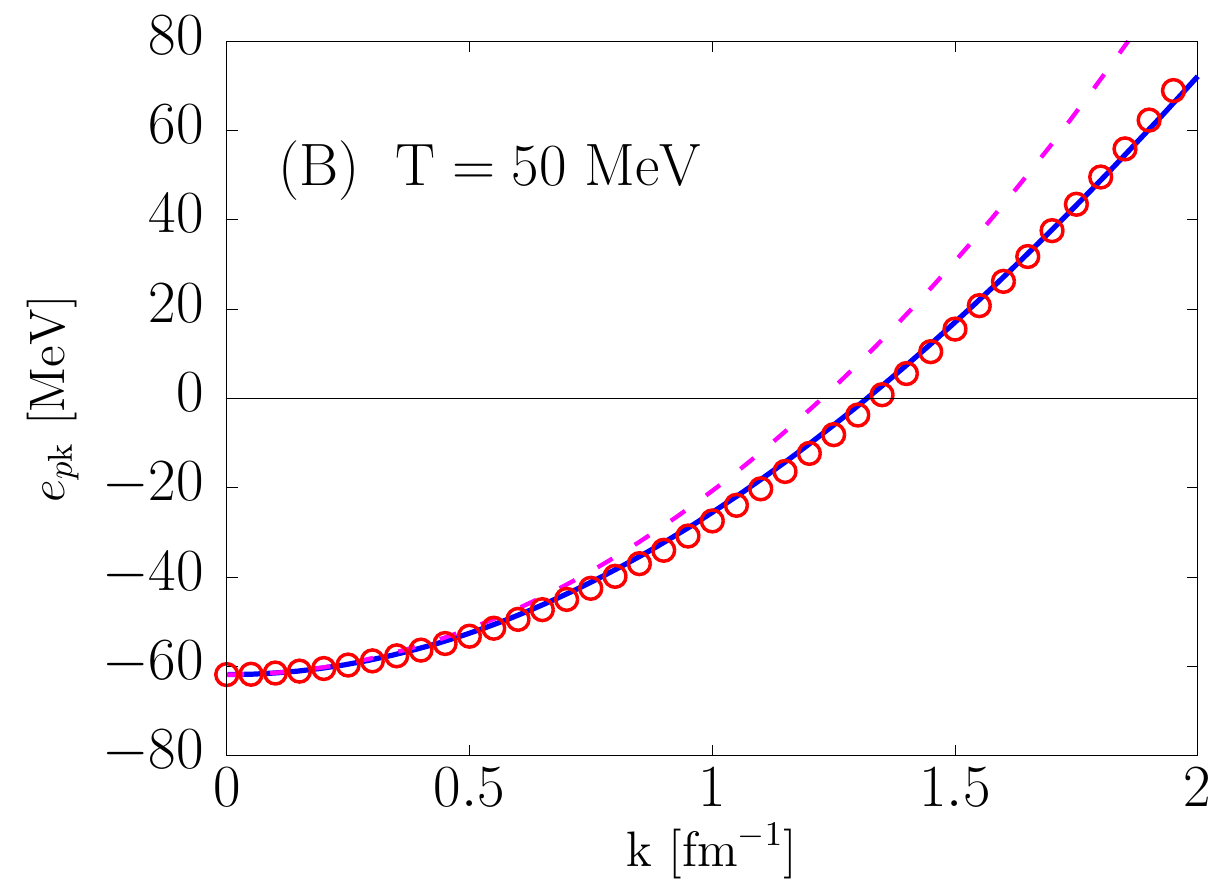}
\caption{Momentum dependence of the proton spectrum in charge-neutral $\beta$-stable matter at baryon density $\varrho = 2 \varrho_0$ and temperature $T=0$ (A) and 50 MeV (B). The solid and dashed lines represent the result of full microscopic calculations and the approximation of 
Eq.~\eqref{ek:quad}, respectively. The open circles in panel (B) have been obtained using the quadratic approximation with  $m^\star_0$  replaced 
by the effective mass computed at $T=50$ MeV.
\label{quadratic:spectra}}
\end{figure}

The solid lines of Fig.~\ref{quadratic:spectra} show the momentum dependence of the proton spectra in charge-neutral $\beta$-stable matter at baryon density $\varrho = 2 \varrho_0$ and temperature $T=0$ and 50 MeV. A comparison with the dashed lines
indicates that at $T= 50$ MeV the apprroximation of Eq.~\eqref{ek:quad} fails to provide an accurate representation of $e_{\alpha k}$ for $k > 0.5 \ {\rm fm}^{-1}$. However, the quadratic approximation turns out to be in remarkable agreement with the microscopic result if  $m^\star_0$  is replaced with the effective mass 
computed at $T=50$ MeV.

\section{Bulk viscosity of neutron star matter}
\label{viscosity}

Bulk viscosity in charge-neutral $\beta$-stable matter appears in the aftermath of an instantaneous departure from $\beta$-equilibrium,  
 resulting from the change of density and pressure induced by some perturbation. The role of bulk viscosity in determining the maximum rotation rate of neutron 
 stars\textemdash the value of which is limited by the onset of the Chandrasekahr-Friedman-Shutz (CFS) instability, driven by gravitational wave emission~\cite{chandra1,chandra2,FS}\textemdash has been discussed by many authors; for a review, see, e.g., 
Ref.~\cite{ak2}. More recent studies, motivated by the observation of the gravitational wave event GW170817~\cite{PhysRevLett.119.161101}, focused on the effect of bulk viscosity in neutron star mergers, leading to a damping of density oscillations~\cite{alfordharris2019}.

\subsection{Dissipative processes in fluids}

The description of fluids is based of the continuity equation expressing conservation of mass
\begin{align}
\label{continuity}
\frac{\partial \varrho}{\partial t} = - \boldsymbol{\nabla} \cdot ( \varrho {\boldsymbol {\rm v}} )  \ , 
\end{align}
where ${\bf v}={\bf v}({\bf r},t)$ and $\varrho = \varrho({\bf r},t)$ denote the velocity field and matter density, respectively.
For ideal fluids, that is, in the absence of dissipation, the force acting on a fluid element is simply related to the pressure of the surrounding medium
through 
\begin{align}
\label{newtonlaw}
{\bf F} = -{\boldsymbol \nabla} P \ , 
\end{align}
and the equation of motion reduces to Euler's equation~\cite{Landau_fluids}
\begin{align}
\frac{\partial {\bold v}}{\partial t}  + ( \bold{ v} \cdot {\boldsymbol \nabla} ) {\bf v}= - \frac{1}{\varrho} {\boldsymbol \nabla} P \ .
\label{euler}
\end{align}
The above equations can be combined to obtain 
\begin{align}
\frac{\partial (\varrho {\rm v}_i)}{\partial t} = - \nabla_j \Pi_{ij}  \ , 
\end{align}
where a sum on the index $j$ is implicit and $\Pi_{ij}$ is the momentum flux tensor, defined as
\begin{align}
\label{momtens1}
\Pi_{ij}  = P \delta_{ij} + \varrho {\rm v}_i {\rm v}_j \ .
\end{align}

In viscous fluids, the form of the continuity equation does not change, but the momentum flux tensor needs to be modified by adding
a term describing the deviations from the ideal fluid behaviour, due the occurrence of processes involving  {\it irreversible} momentum transfer. The resulting expression can be written in the form
\begin{align}
\label{momtens2}
\Pi_{ij}  = 
 P \delta_{ij} + \varrho {\rm v}_i {\rm v}_j + \delta \Pi_{ij}  \ , 
\end{align}
where
\begin{align}
\delta \Pi_{ij} = - \eta  \Big[  \nabla_j {\rm v}_i +  \nabla_i {\rm v}_j  + \frac{2}{3} \delta_{ij} ( {\boldsymbol \nabla} \cdot {\bf v} )  \Big]
- \zeta  \delta_{ij} ( {\boldsymbol \nabla} \cdot {\bf v} )  \ ,
\end{align}
with the two velocity-independent quantities $\eta$ and $\zeta$ being referred to as shear and bulk viscosity coefficients, respectively.
In the case of incompressible fluids, in which $\varrho$ does not depend on either ${\bf r}$ or $t$, Eq.~\eqref{continuity}  implies 
\begin{align}
\boldsymbol{\nabla} \cdot {\boldsymbol {\rm v}}  = 0  \ , 
\end{align}
and the contribution of bulk viscosity vanishes.

A pulsation of frequency $\omega$ induces a variation of the fluid density described by the equation
\begin{align}
\label{dens:fluct}
\varrho(t) = \varrho_{\rm eq} + \delta \varrho \cos \omega t \ , 
\end{align}
where $\varrho_{\rm eq}$ is the density corresponding to chemical equilibrium and  $( \delta \varrho/ \varrho_{\rm eq} ) \ll 1$. 
The rate of energy-density dissipation due to bulk viscosity, averaged over the pulsation period $\tau = 2 \pi/\omega$, is given 
by~\cite{Schaefer:2014awa}
\begin{align}
\label{edot:fluid}
\left\langle \frac{d \epsilon_{\rm diss} }{dt} \right\rangle  = - \frac{1}{\tau} \int_0^\tau  dt \ 
\nabla_i   \left[   - v_j \ \zeta  \delta_{ij} ( {\boldsymbol \nabla} \cdot {\bf v} ) \right]  = 
 \frac{1}{\tau} \int_0^\tau  dt \  \zeta ( {\boldsymbol \nabla} \cdot {\bf v} )^2 \ .
\end{align}

\subsection{Bulk viscosity of $\beta$-stable matter}
\label{bulk:beta}

The authors of  refs.~\cite{haenselschaeffer1992,haenseletal2000} have derived the expression of the bulk viscosity coefficient of $\beta$-stable 
matter under the assumption that neutrinos and antineutrinos are produced through the processes 
\begin{align}
\label{eq:dUrca2}
    p + e & \to n + \nu_e \ ,  \\
\label{eq:dUrca1}    
    n & \to p + e + {\overline \nu}_e  \ , 
\end{align}
referred to as {\it direct} Urca reactions. They followed the scheme originally proposed in Ref.~\cite{Sawyer}, and employed a simple parametrisation of the nuclear matter EOS assuming a quadratic dependence on the neutron excess $\alpha = 1-2Y_p$. 

Bulk viscosity is associated with deviations from beta equilibrium, signaled by a non vanishing difference between the 
neutrino and antineutrino production rates
\begin{align}
\Delta \Gamma = \Gamma_{\nu_e} - \Gamma_{ {\overline \nu}_e } \  ,
\end{align}
with $\Delta \Gamma$ being a function of the variable $\delta \mu$, describing the departure from chemical equilibrium.  In the
case of non degenerate neutrinos
\begin{align}
\delta \mu = \mu_n - \mu_p - \mu_e \ . 
\end{align}
Under the assumption that $\delta \mu \ll T \ll \mu_i$, $\Delta \Gamma$ can be expanded in powers of $\delta_\mu$. At leading 
order one finds
\begin{align}
\Delta \Gamma = \lambda \delta_\mu  \  , 
\end{align}
with
\begin{align}
\label{def:lambda}
\lambda = 2 \Big(  \frac { \partial \Gamma_\nu }{ \partial  \delta \mu } \Big)_{\delta \mu = 0} \ .
\end{align}

The energy-density dissipation rate\textemdash averaged over the period of the density fluctuation of Eq.~\eqref{dens:fluct}\textemdash can be written in terms of the change of pressure associated with  the departure from chemical equilibrium using
\begin{align}
\nonumber
\left\langle \frac{ d \epsilon_{\rm diss} }{ d t } \right\rangle & =  
\frac{1}{\varrho}~\frac{1}{\tau}~\int_0^\tau~dt~\delta P_{\rm chem}~\frac{d \varrho}{dt} \\ 
\label{edot0:npe}
& = \frac{\delta \varrho}{\varrho}~\frac{\omega}{\tau}~\int_0^\tau~dt~\delta P_{\rm chem}~\sin \omega t \ , 
\end{align}
with~\cite{Sawyer}
\begin{align}
\label{def:deltaP}
\delta P_{\rm chem} = - \lambda C^2~\frac{\delta \varrho}{\varrho}~\frac{\omega } 
{ \omega^2 + (2 \lambda B/\varrho)^2 }~\sin \omega t + \ldots
\end{align}
where the ellipses refer to additional terms giving vanishing contributions to the integral. Here,  $\lambda$ is given by Eq.~\eqref{def:lambda}, and the constants $B$ and $C$ are defined as 
\begin{align}
\label{def:BC}
B = \Big(  \frac{ \partial \delta \mu} { \partial \alpha }\Big)_{\delta \varrho = 0} \ \ \ , 
\ \ \ C = \Big[ \varrho~\Big(  \frac{ \partial \delta \mu} { \partial \varrho }\Big) \Big]_{\alpha = \alpha_{\rm eq}}  \ ,
\end{align}
where $\alpha_{\rm eq}$ is the value of neutron excess corresponding to $\beta$-equilibrium.
Substitution of Eq.~\eqref{def:deltaP} into Eq.~\eqref{edot0:npe} yields
\begin{align}
\nonumber
\label{edot:npe}
\left\langle \frac{d \epsilon_{\rm diss} }{dt} \right\rangle  = - \lambda~\frac{\omega^2}{2}~\Big( \frac{ \delta \varrho}{\varrho } \Big)^2~\frac{C^2}{ \omega^2 + (2 B \lambda/\varrho)^2 } \ . 
\end{align}
The bulk viscosity coefficient is obtained combining the above equation with Eqs.~\eqref{edot:fluid}, which can be rewritten using the 
continuity equation associated with conservation of baryon number, implying 
\begin{align}
{\boldsymbol \nabla} \cdot  {\bf v}  = - \frac{1}{\varrho}~\frac{ \partial \varrho }{ \partial t } 
= \omega~\frac{\delta \varrho}{\varrho}~\sin \omega t \ .
\end{align}
Following this procedure, one finally arrives at 
\begin{align}
\label{zeta:old}
\zeta = -\lambda~\frac{C^2}{ \omega^2 + (2 B \lambda/\varrho)^2 } \ .
\end{align}
The above result shows that the calculation of the bulk viscosity coefficient involves $\lambda$, which is
obtained from the neutrino and antineutrino production rates, and the constants $B$ and $C$ defined 
by Eqs.~\eqref{def:BC}. 
We have studied the density and temperature dependence of $\zeta$ using 
the nuclear matter model described in the previous sections, which allows to take into account 
thermal modifications of the chemical potentials and effective masses within a fully consistent framework.

\subsection{Calculation of the bulk viscosity coefficient}
\label{calc:zeta}

The calculations of the rates of neutrino and antineutrino production in the Urca processes of Eqs.~\eqref{eq:dUrca2} and~\eqref{eq:dUrca1}\textemdash needed to obtain $\lambda$ from Eq.~\eqref{def:lambda}\textemdash  have been performed using an {\it improved} version of the commonly used Fermi surface approximation. The  
important new feature of this procedure lies in the fact that   
the effective masses depend on temperature, and so does the phase space integration.

Equation~\eqref{zeta:old} can be written in a somewhat more transparent form in terms of the equilibration 
rate $\gamma = - 2 B \lambda /\varrho$. The resulting expression
\begin{align}
\zeta = \varrho \frac{C^2}{B}~\frac{1}{2}~\frac{\gamma}{ \omega^2 + \gamma^2 } \ ,
\end{align}
exhibits a resonant maximum located at $\gamma = \omega$. This property, which depends on both density and temperature,  has been thoroughly analysed by the authors of Ref.~\citep{alfordharris2019}, who also discussed the differences between the  isothermal and adiabatic treatment of thermodynamic quantities. 

Under the assumption that during a neutron star merger the heat flow
between adjoining fluid elements be negligible, the derivatives involved in the calculations of
the quantities $B$ and $C$ defined by Eq.~\eqref{def:BC} must be computed keeping the entropy per baryon constant. 
The alternative procedure  based on isothermal derivation, while being equivalent in the $T \to 0$ limit,  
leads to sizeably different  results in the high-temperature regime. 
\begin{figure}[h]
\includegraphics[scale=0.70]{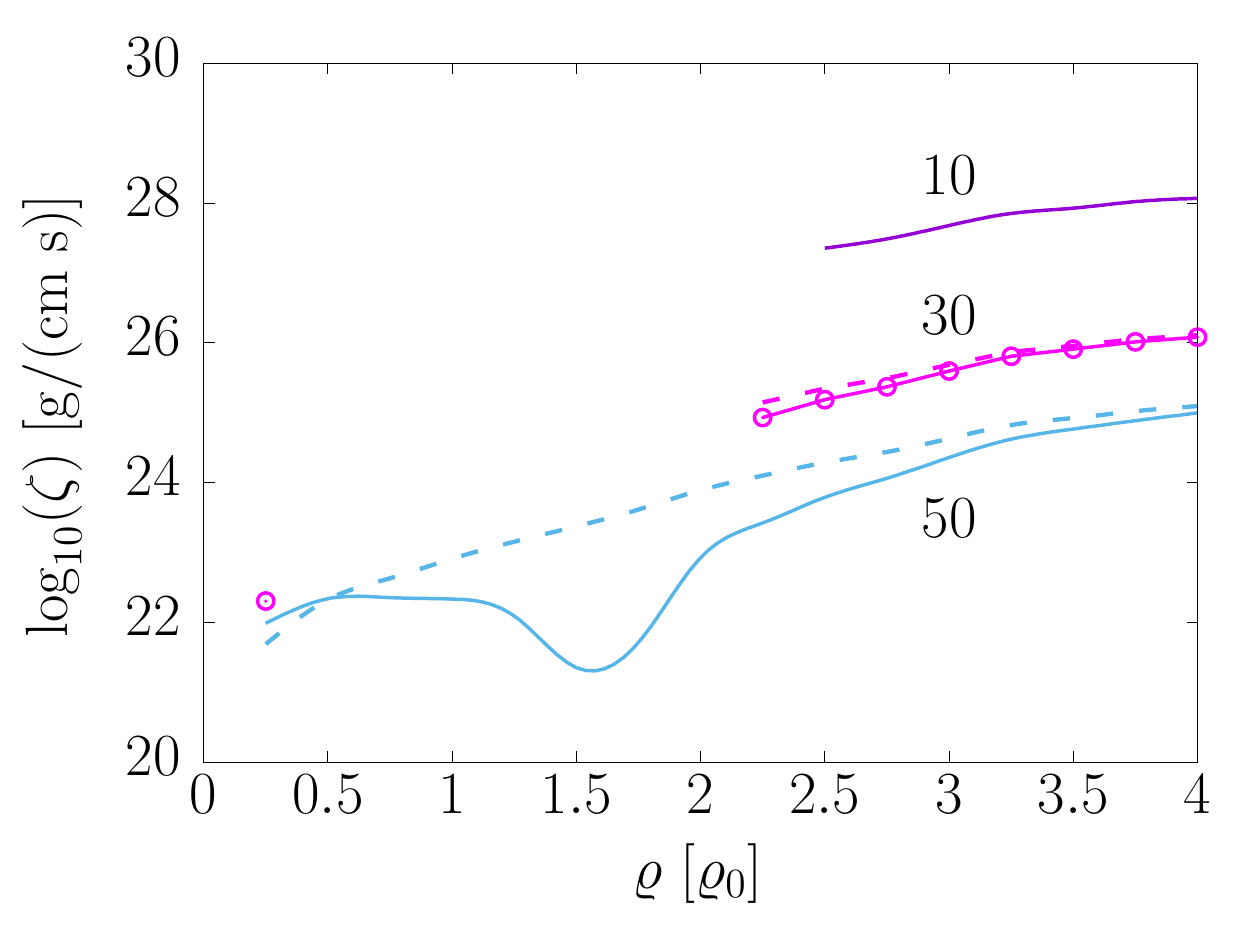}
\vspace{-0.25cm}
     \caption{Density dependence of the bulk viscosity coefficient of $\beta$-stable matter associated with a density fluctuation of frequency  $\omega = 2\pi \times 1 \ \mathrm{kHz}$. The results have been obtained by performing the derivatives of Eqs.~\eqref{def:BC} using both the isothermal (solid lines and open circles) and adiabatic (dashed lines) definitions.The labels specify the temperature in units of MeV.}
\label{fig:BV_dens_highT}
\end{figure}

Figure~\ref{fig:BV_dens_highT} shows the behaviour of $\zeta$ as a function of 
density for temperatures 10, 30, and 50 MeV, with the frequency of the density oscillation driving the appearance of viscosity being set to $\omega = 2\pi \times 1 \ \mathrm{kHz}$, a value typical of neutron star pulsations. 
The results have been obtained by performing the derivatives of Eqs.~\eqref{def:BC} using both the isothermal (solid lines and open circles) and adiabatic (dashed lines) definitions. 

The peculiar density dependence featured by the solid line
corresponding to the highest temperature, $T=~50$~MeV, results from  the occurrence of a minimum in the proton fraction $Y_p(\varrho)$\textemdash clearly 
visible in Fig.~\ref{prot:frac}\textemdash and from the fact that the value of $Y_p$ exceeds the threshold for the onset of direct Urca processes at all densities. A comparison to the corresponding dashed line shows that the minimum of $\zeta(\varrho)$ is entirely   washed out when the isothermal derivative is replaced by the adiabatic one. Because the proton fraction is still above the Urca threshold, however,  one finds $\zeta(\varrho)\neq0$ over the whole density range of the figure.    

A similar pattern is displayed by the solid line and the open circles representing the results obtained at $T=30 \ \mathrm{MeV}$, although in this case 
the minimum of $Y_p(\varrho)$ is not reflected by a minimum in $\zeta(\varrho)$. 
One only observes an isolated point with $\zeta \neq 0$ located at $\rho = 0.25~\rho_0$, and
a density region extending  up to $\rho > 2.25~\varrho_0$ in which $\zeta = 0$. At  larger densities, $Y_p$ is 
always above the threshold of Urca processes, and $\zeta \neq 0$. At $T=10$ MeV the solid and dashed lines turn out to lie on top of one another.

It should be noted that, although at high temperatures the Urca process is always allowed, the corresponding values 
of $\zeta$ are very small, and  bulk viscosity is unlikely to be distinguished from other dissipative phenomena active in
 a neutron-star merger.

\begin{figure}[htb]
\includegraphics[scale=0.70]{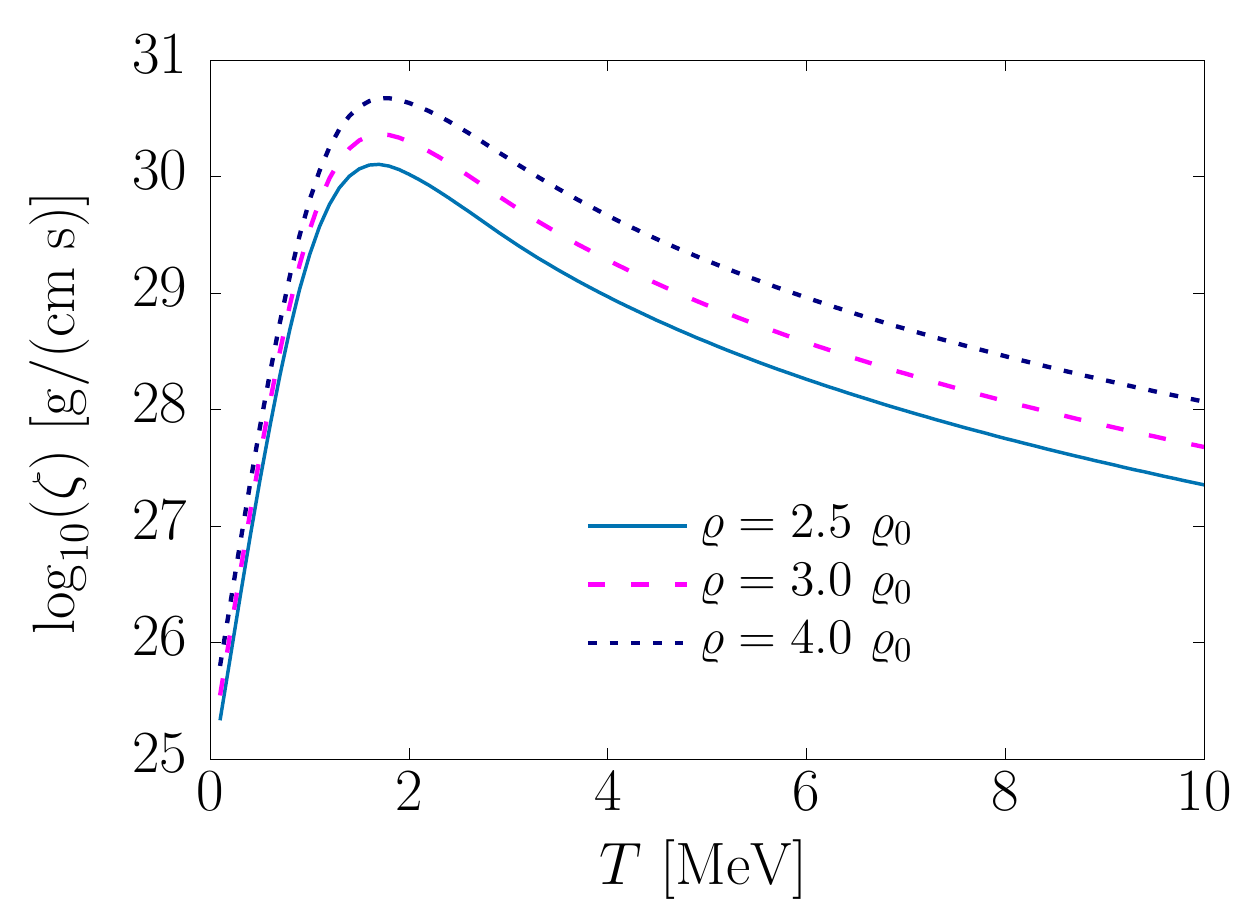}
\vspace{-0.25cm}
     \caption{Temperature dependence of the bulk viscosity coefficient $\zeta$  corresponding to 
     $\omega~=~2\pi \times~1 \ \mathrm{kHz}$ and different densities.}
\label{fig:BV_temp}
\end{figure}

The temperature-dependence of the bulk viscosity coefficient at $\omega = 2\pi \times 1 \ \mathrm{kHz}$ and densities $\varrho =$ 2.5, 3, and 4 $\varrho_0$ is illustrated in  Fig.~\ref{fig:BV_temp}. The maximum at $T\approx2 \ \mathrm{MeV}$, 
whose value monotonically increases with density, is clearly apparent. Owing to the low-temperature range 
spanned by the figure, in this instance the results obtained using isothermal and adiabatic derivatives 
turn out to be nearly identical. For this reason, only isothermal results are shown.

\section{Discussion}

The description of neutron-star mergers, which will be needed for the interpretation of current and future astronomical observations, requires a quantitative understanding of both equilibrium and dynamical properties of hot and dense matter. In this context, the availability of a dynamical model strongly constrained by phenomenology and suitable for use in finite-temperature perturbation theory will be crucial.

The approach described in this paper has been employed to obtain the EOS of matter with arbitrary neutron excess 
in the density region extending 
up to 4~$\varrho_0$\textemdash in which the applicability of the description in terms of nucleons is supported by 
electron-nucleus scattering data~\cite{benhar:yscaling}\textemdash and temperatures up to 50 MeV. Single-nucleon
properties, such as the quasipartlcle spectra and effective masses, have been also computed within a theoretical  
framework in which thermal effects are consistently taken into account. Exploratory studies of dissipative processes
leading to the damping of density oscillations in neutron-star mergers suggest that a detailed treatment of thermal 
effects is needed to clarify important issues, such as the onset of the Urca process.

The extension of the approach outlined in this paper to the description of neutrino emission processes and the neutrino mean free path along the line described in Refs.~\cite{Lovato:2012ux,LBGL} does not involve any conceptual issues, 
and is currently under way.   





\vspace{6pt} 



\authorcontributions{All authors contributed to conceptualisation, development of the formalism and the computing codes, and manuscript preparation. All authors have read and agreed to the published version of the manuscript.
}

\funding {This research was funded by the U.S. Department of Energy, Office of Science, Office of Nuclear Physics, under contract DE-AC02-06CH11357 (A.L.), the NUCLEI SciDAC program (A.L.), and the Italian National Institute for Nuclear Research (INFN), under grant TEONGRAV (O.B. and L.T.).}

\dataavailability{The data supporting the conclusions of this study are available in the cited literature and within the present article.} 


\conflictsofinterest{The authors declare no conflict of interest.} 




\begin{adjustwidth}{-\extralength}{0cm}

\reftitle{References}

\PublishersNote{}
\end{adjustwidth}
\end{document}